\documentclass[10pt,a4paper]{article}
\usepackage{graphics}
\usepackage{latexsym}
\usepackage{amssymb}
\usepackage{bm}
\usepackage{cite}
\begin{document}

\title{Evaluation of bistable systems versus matched filters in detecting
bipolar pulse signals}

\author{Fabing Duan (1), Derek Abbott (2), Qisheng Gao (1)\\
\small (1) Institute of Complexity Science, Department of
Automation Engineering, Qingdao University, \\ \small Qingdao
266071, People's Republic of China\\ \small (2) Centre for
Biomedical Engineering
(CBME) and School of Electrical \& Electronic Engineering, \\
\small The University of Adelaide, Adelaide, SA 5005, Australia}
\maketitle

\begin{abstract}
This paper presents a thorough evaluation of a bistable system
versus a matched filter in detecting bipolar pulse signals. The
detectability of the bistable system can be optimized by adding
noise, i.e.~the stochastic resonance (SR) phenomenon. This SR
effect is also demonstrated by approximate statistical detection
theory of the bistable system and corresponding numerical
simulations. Furthermore, the performance comparison results
between the bistable system and the matched filter show that (a)
the bistable system is more robust than the matched filter in
detecting signals with disturbed pulse rates, and (b) the bistable
system approaches the performance of the matched filter in
detecting unknown arrival times of received signals, with an
especially better computational efficiency. These significant
results verify the potential applicability of the bistable system
in signal detection field.
\end{abstract}

\section{Introduction}

Stochastic resonance (SR) phenomena involve optimizing a system
performance measure via adding noise, and have received
considerable attention in the past few decades
\cite{Benzi1,Benzi2,Ganopolski,McNamara1989,Gammaitoni1998,Wiesenfeld1995,Moss,Bulsara,Morse,Russell,Glass,Anishchenko1999,Hanggi,Allison,Fauve,McNamara1988,Spano,Gammaitoni1991,Hibbs,Barbay,Apostolico,Duan2004}.
The prototype SR model is an overdamped bistable system, which was
used to describe the earth's climatic change
\cite{Benzi1,Benzi2,Ganopolski}. Since then, this generic system
is a frequently used model for characterizing SR phenomena in
diverse scientific fields
\cite{McNamara1989,Gammaitoni1998,Wiesenfeld1995,Moss,Bulsara,Morse,Russell,Glass,Anishchenko1999,Hanggi,Allison}.
Moreover, many experimental verifications of SR effects have been
demonstrated for a bistable characteristic, such as in a Schmitt
Trigger \cite{Fauve}, bistable ring laser \cite{McNamara1988},
paramagnetically driven bistable buckling ribbon \cite{Spano},
bistable electron paramagnetic resonance systems
\cite{Gammaitoni1991}, bistable superconducting quantum
interference devices \cite{Hibbs}, vertical cavity surface
emitting lasers \cite{Barbay}, etc. In addition, new SR-type
phenomena were also observed in bistable systems subjected to
suprathreshold input signals \cite{Apostolico,Duan2004}.

Recent research has focused on potential SR applications in signal
transmission
\cite{Barbay,Collins,Chapeau-Blondeau1997,Godivier,Kish,Duan},
estimation \cite{Papadopoulos,Chapeau-Blondeau2002} and detection
\cite{Inchiosa,Bulsara&Zador,Galdi,Inchiosa2,Kay,Chapeau-Blondeau2000,Kosko,Jung,Ando,Asdi,Luchinsky,Harmer1,Harmer2,Zozor,Zozor2003,Tougaard2000,Tougaard2002,Bulsara2003,Harmer}.
Signal detection theory offers a powerful tool for analyzing
linear systems, such as the matched filter with a Gaussian white
noise background and its impulse response matched to the input
signal maximizing the filter output signal-to-noise ratio
\cite{Kay1998}. These nonlinear systems, thereby, are potentially
rich in signal detection applications, though generally more
difficult to theoretically tackle
\cite{Gammaitoni1998,Moss,Kay,Chapeau-Blondeau2000}. A variety of
appealing nonlinear systems or models exhibiting SR effects have
been explored as detection devices, such as the threshold detector
\cite{Chapeau-Blondeau2000,Jung}, electronic circuits
\cite{Gammaitoni1998,Luchinsky,Harmer1,Harmer2}, biological
systems \cite{Gammaitoni1998,Wiesenfeld1995,Moss,Russell,Hanggi},
neuron models \cite{Tougaard2000,Tougaard2002,Harmer}, bistable
systems \cite{Inchiosa,Bulsara&Zador,Galdi}, etc. Although many
researchers stressed that SR can not do wonders in signal
detection, in the sense that nonlinear systems exploiting SR can
never do better than the corresponding linear systems
\cite{Inchiosa,Bulsara&Zador,Galdi,Tougaard2002,Dykman,Peterson,Peterson&Gebeshuber,McDonnell}
in the presence of Gaussian white noise, some promising results
have been obtained for non-Gaussian noise conditions
\cite{Kay,Chapeau-Blondeau2000,Kosko,Zozor,Zozor2003}, the
computational and design efficiency of nonlinear systems
\cite{Galdi,Papadopoulos1998}, noise-floor limited systems
\cite{Inchiosa2000,Nikitin}, psychophysical experiments
\cite{Collins} or models \cite{Gong}, and parallel arrays of
nonlinear devices via the suprathreshold SR effect
\cite{Stocks2001,Stocks2000,McDonnell1,McDonnell2,David}.

In line with this, we investigate further the performance of
bistable systems versus matched filters in detecting bipolar pulse
signals. Some new merits of bistable systems will be emphasized in
the comparison of results. Section~\ref{sec:II} develops a
theoretical non-stationary probability density model of the
bistable system. This approximate theory describes the SR
phenomenon in the bistable system well, and agrees with the
numerical simulation results. The comparison of the pertinent
signal detection statistics, between bistable systems and matched
filters, demonstrates the fact that the bistable system is a
suboptimal detector against the matched filter
\cite{Inchiosa,Bulsara&Zador,Galdi,Stocks2001,Stocks2000}. Beyond
this fact, we evaluate in detail the detectability of bistable
systems, versus matched filters, in detecting pulse signals with
disturbed pulse rates and unknown arrival times in
Sec.~\ref{sec:III}. From the power spectrum analysis of bistable
system outputs, the different pulse rates of the disturbed signal
can be clearly discriminated. The matched filter, however, is more
like a lowpass filter, wherein too much noise at low frequency
bands weakens its resolution. In the case of detecting the unknown
arrival time of received signals, the bistable system, with lower
computational overhead, is an appropriate substitute for the
matched filter. The corresponding numerical simulation results
also indicate that the bistable system is a more competitive
detector than the matched filter, with respect to its robustness
and computational efficiency. This may be of importance to
bistable electronic or optical devices utilized in signal
detection. We argue that the potential applicability of the
bistable system deserves further study in other signal processing
problems.


\section{\label{sec:II} Signal detection statistics
of the bistable system versus the matched filter}

Here, we consider the detection problem of a bipolar pulse signal
$s(t)$ and introduce two hypotheses
\begin{eqnarray}
\textit{\textbf{H}}_{0}: s(t)=-A,\: [(n-1)T_p,\: nT_p],\nonumber \\
\textit{\textbf{H}}_{1}: s(t)=+A,\: [(n-1)T_p,\: nT_p],
\label{eq:one}
\end{eqnarray}
where $n=1,2,\ldots$. The input pulse signal $s(t)$, for
$(n-1)T_p\leq t\leq nT_p$, has the pulse duration time $T_{p}$ and
pulse amplitudes $\pm A$ (see an example of Fig.~\ref{fig:one}).
The background noise $\eta(t)$ is additive Gaussian white noise
with autocorrelation $\left\langle \eta(t)\eta(0)\right\rangle
=2D\delta (t)$ and zero-mean. Here, $D$ denotes the noise
intensity. The mixture of $s(t)$ plus $\eta(t)$ is then applied to
a detector, such as the matched filter or the bistable system
introduced below.

\subsection{Nonstationary probability density model of the bistable system}
The bistable system under study is described as
\begin{equation}
\tau_{a}\frac{dx(t)}{dt}=x(t)-\frac{x^{3}(t)}{X_{b}^{2}}+s(t)+\eta(t),\label{eq:two}
\end{equation}
with real system parameters $\tau_a$ and $X_b$ \cite{Godivier}.
$\tau_a$ is related to the system relaxation time. The dynamics of
Eq.~(\ref{eq:two}) are derived from the symmetrical double-well
potential $V_0(x)=-x^2/2+x^4/(4X_b^2)$, having the two minima
$V_0(\pm X_b)=-X_b^2/4$. Parameters $\tau_a$ and $X_b$ have the
units of time and signal amplitude respectively, and define
natural scales associated to the process of Eq.~(\ref{eq:two}).

In each pulse duration $T_p$, the system of Eq.~(\ref{eq:two}) is
subjected to the constant amplitude $+A$ or $-A$, with an additive
input Gaussian white noise $\eta(t)$. Under these conditions, the
statistically equivalent description for the corresponding
probability density $\rho(x,t)$ is governed by the Fokker-Planck
equation
\begin{equation}
\tau_a \frac{\partial \rho(x,t)}{\partial
t}=\big[\frac{\partial}{\partial x} V'(x)+
\frac{D}{\tau_a}\frac{\partial^2}{\partial x^2}\big]\rho(x,t),
\label{eq:three}
\end{equation}
where $V'(x)=-x+x^3/X_b^2\mp A$ and the Fokker-Planck operator is
$L_{FP}=\frac{\partial}{\partial x}
V'(x)+\frac{D}{\tau_a}\frac{\partial^2}{\partial x^2}$
\cite{Risken}. Here, $\rho(x,t)$ obeys the natural boundary
condition that it vanishes at large $x$ for any $t$. The
steady-state solution of Eq.~(\ref{eq:three}), for a permanent
input at $+A$ or $-A$, is given by
\begin{equation}
\rho(x)=\lim_{t\rightarrow\infty}\rho(x,t)=C\exp[-\frac{\tau_a
V(x)}{D}], \label{eq:four}
\end{equation}
where $C$ is the normalization constant \cite{Risken}. This
solution of Eq.~(\ref{eq:four}) is for global equilibrium in the
double-well potential.

Now, we will seek the nonstationary solution $\rho(x,t)$ of the
Fokker-Planck equation, Eq.~(\ref{eq:three}), in the case of an
input transition from $s(t)=-A$ to $s(t)=+A$, or vice versa. This
calculation is carried out in Appendix~A. We show in Appendix~A
that the transition from the stationary density corresponding to
$s(t)=-A$ to the stationary density corresponding to $s(t)=+A$, or
vice versa, is dominated by an exponential temporal relaxation
with a time constant $1/\lambda_1$. This allows us to deduce a
response time $T_r=\tau_a/\lambda_1$ for the bistable system,
which is a measure of the time taken by the system to switch from
one potential well to the other, when the input changes from $\pm
A$ to $\mp A$, in the presence of noise. In Appendix~A, the
non-stationary solution $\rho(x,t)$ for an input transition from
$s(t)=-A$ to $s(t)=+A$ (or vice versa) is approximated with the
two first terms from its asymptotic representation of
Eq.~(\ref{eq:A11}), as
\begin{equation}
\rho(x,t|s(t)=\pm A)\simeq \rho(x|s(t)=\pm A) +[\rho(x|s(t)=\mp
A)-\rho(x|s(t)=\pm A)]\exp(-t/T_r), \label{eq:five}
\end{equation}
where $\rho(x|s(t)=\pm A)$ are the steady-state solutions of
Eq.~(\ref{eq:four}). In Eq.~(\ref{eq:five}), when $t=0$, the term
$\exp(-t/T_r)=1$ and $\rho(x,t|s(t)=\pm A)$ starts with the
initial condition of $\rho(x|s(t)=\mp A)$. As
$t\rightarrow+\infty$, the term $\exp(-t/T_r)=0$, and
$\rho(x,t|s(t)=\pm A)$ tends to the stationary condition of
$\rho(x|s(t)=\pm A)$. This theoretical model, although
approximate, nicely captures the double role played by the noise,
both in accelerating the switching dynamics between wells and in
enhancing the fluctuations inside the wells. With this
nonstationary solution of Eq.~(\ref{eq:five}), as we will see, the
nonmonotonic evolution of the signal detection statistics of the
bistable system can be analyzed theoretically.

\subsection{Signal detection statistics of the bistable system and the matched filter}
We are interested in detecting the input pulse signal $s(t)$ from
the observation of the outputs of bistable system or matched
filter. The input pulse signal $s(t)$ is emitted at a rate of one
waveform every $T_p$ and lasts over a pulse duration $T_p$. In
this detection problem, we assume that the interval $T_p$ is both
known at the emitter and the detector. But, the waveform
amplitudes $+A$ and $-A$ are not known at the detector. In order
to detect $s(t)=+A$ or $s(t)=-A$, we sample the system output
signals at equispaced times $t_j=jT_P$ for $j=1,2,\ldots$. The
probability density $\rho(x,T_p|s(t)=\pm A)$ of
Eq.~(\ref{eq:five}) is then easily obtained for the bistable
system. For the input bipolar pulse signals $s(t)$ mixed with
additional Gaussian white noise $\eta(t)$, the impulse response of
matched filter is $h(t)=s(T_{p}-t)$, where $h(t)$ is a scaled,
time-reversed and shifted version of the pulse signal $s(t)$
($0<t\leq T_p$). If the output of the matched filter is sampled at
$jT_p$ that maximizes the output signal-to-noise ratio, the
probability density $\rho(x,T_p|s(t)=\pm A)$ has a Gaussian
distribution
\begin{equation}
\rho(x,T_p|s(t)=\pm
A)=\frac{1}{\sqrt{2\pi}\sigma}\exp[-\frac{(x\mp
A)^{2}}{2\sigma^{2}}],\label{eq:six}
\end{equation}
with $\sigma=\sqrt{A^2 T_P/2D}$ \cite{Kay1998}.

\begin{figure}
\centering{\resizebox{8cm}{!}{\includegraphics{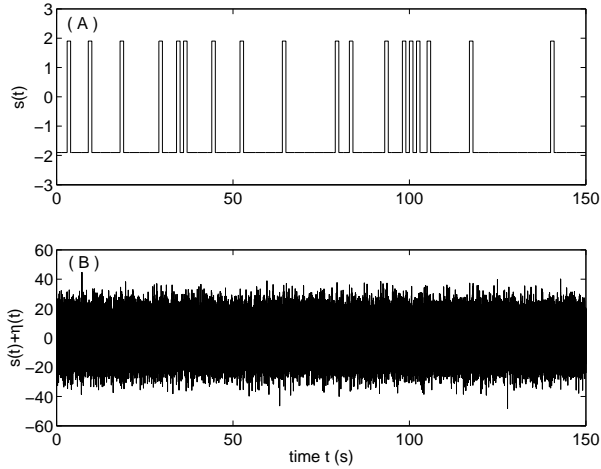}}}
\caption{Time evolution of the signals for the system of
Eq.~(\ref{eq:two}) with $\tau_a=0.1$ s and $X_b=5$ V. The sampling
time step $\Delta t=0.001$ s. (A) The input pulse signal $s(t)$ is
with $A=1.9$ V and $T_p=1$ s. $s(t)=+A$ appears randomly in
successive time intervals $[(n-1)T_p,\: nT_p]$ ($n=1,2,\ldots$);
(B) The mixture of signal plus noise ($D=0.1$ V$^2$/Hz).
\label{fig:one} }
\end{figure}

In this paper, the signal detection statistics of the bistable
system and the matched filter are measured by the false alarm
probability
\begin{equation}
P_{FA}=\int_{\ell}^{+\infty}
\rho(x,T_p|s(t)=-A)dx,\label{eq:seven}
\end{equation}
and the detection probability
\begin{equation}
P_{D}=\int_{\ell}^{+\infty} \rho(x,T_p|s(t)=+A)dx,\label{eq:eight}
\end{equation}
where $\ell$ is the detection threshold, and the positive pulse
signal $s(t)=+A$, shown in Fig.~\ref{fig:one}, is considered as
the information carrying part of the signal in this detection
problem.

\begin{figure}
\centering{\resizebox{6.2cm}{!}{\includegraphics{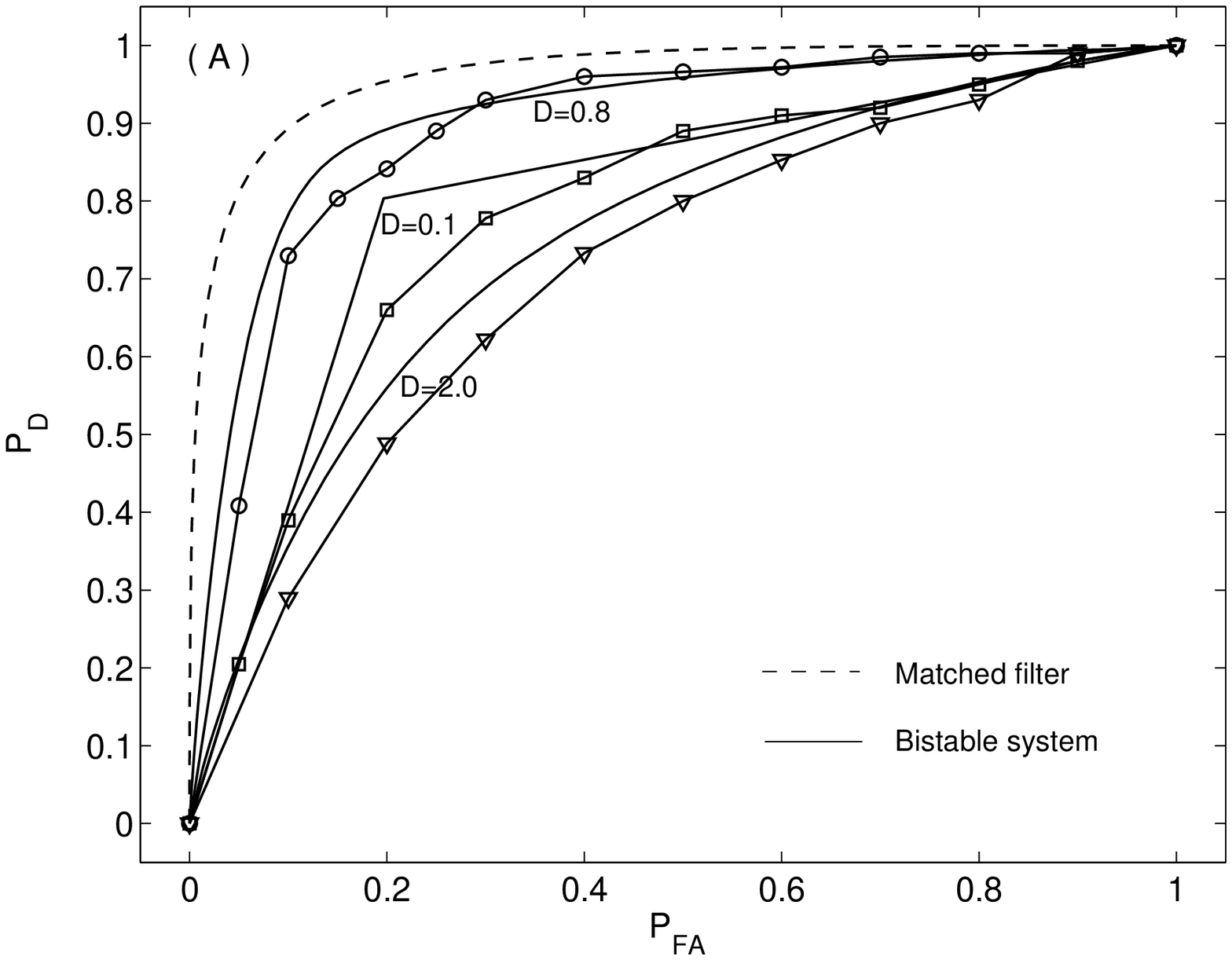}}}
\centering{\resizebox{6.2cm}{!}{\includegraphics{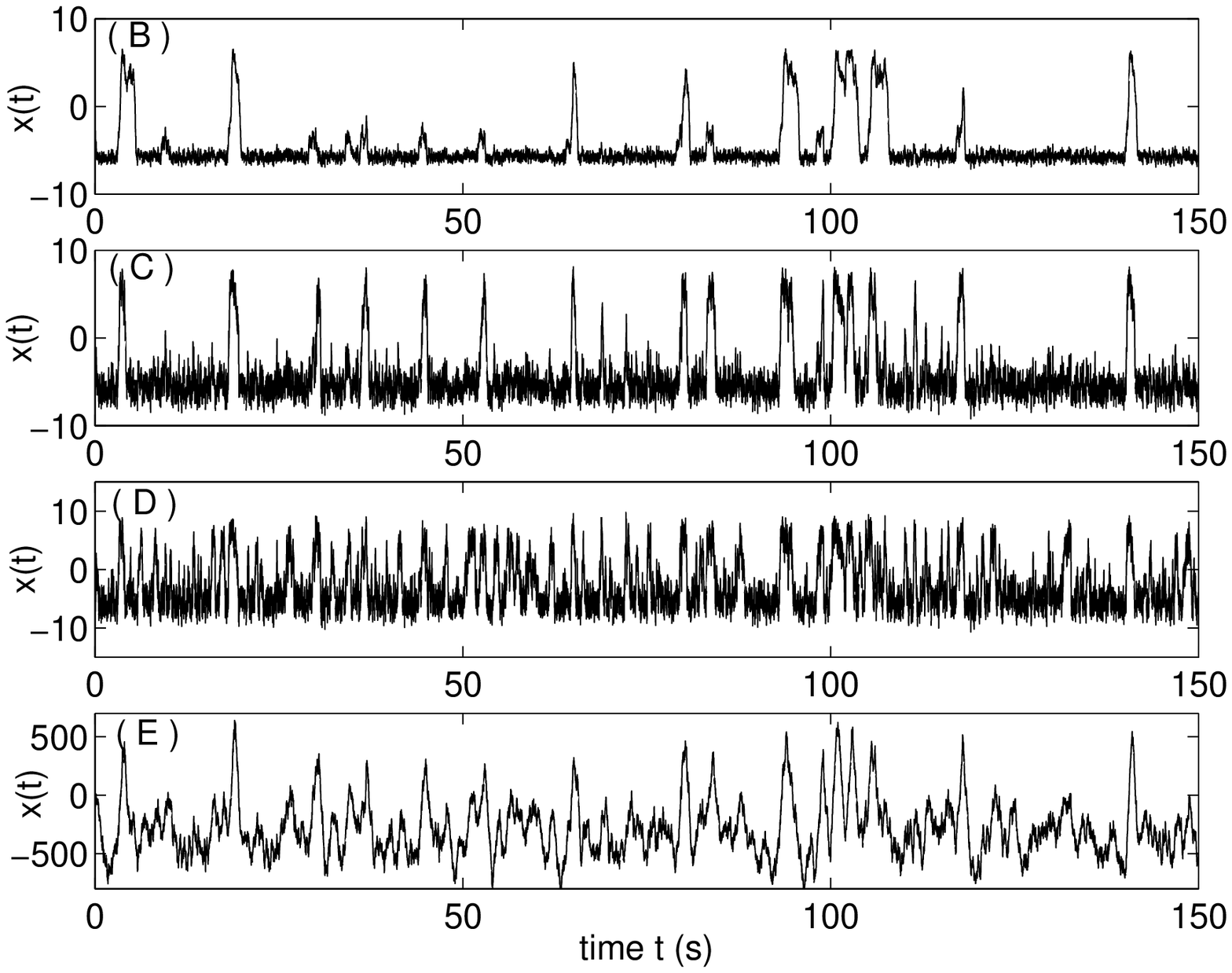}}}
\caption{(A) The detection statistics of the bistable system
(solid lines) with $\tau_a=0.1$ s and $X_b=5$ V at $D=0.8$
V$^2$/Hz, $0.1$ V$^2$/Hz and $2.0$ V$^2$/Hz from the top down, and
the matched filter (dashed line) at $D=0.8$ V$^2$/Hz. Numerical
results of the bistable system are also presented at $D=0.8$
V$^2$/Hz (circles), $D=0.1$ V$^2$/Hz (squares) and $D=2.0$
V$^2$/Hz (down triangles), respectively. The output signals $x(t)$
of the bistable system at (B) $D=0.1$ V$^2$/Hz, (C) $D=0.8$
V$^2$/Hz and (D) $D=2.0$ V$^2$/Hz; (E) The output signal $x(t)$ of
the matched filter at $D=0.8$ V$^2$/Hz. $\Delta t=0.001$ s,
$A=1.9$ V and $T_p=1$ s. \label{fig:two} }
\end{figure}

Figure~\ref{fig:two} (A) compares the signal detection performance
of the bistable system and the matched filter. As the noise
intensity $D$ increases from $0.1$ V$^2$/Hz to $2.0$ V$^2$/Hz, the
signal detection statistics of the bistable system displays a
non-monotonic behavior, i.e.~the SR phenomenon in signal
detection. The detectability of the bistable system is optimized
at $D=0.8$ V$^2$/Hz. The discrete points, illustrated in
Fig.~\ref{fig:two} (A), represent the corresponding numerical
simulation results of the bistable system. Here, we numerically
integrate the stochastic differential equation of
Eq.~(\ref{eq:two}) using a Euler-Maruyama discretization method
with a small sampling time step $\Delta t\ll\tau_a$ \cite{Gard}.
The theoretical model of Eq.~(\ref{eq:five}), although
approximate, clearly describes the SR effect in the bistable
system, and it agrees with the corresponding numerical results.
Furthermore, the output signals of the bistable system and the
matched filter are also given in Figs.~\ref{fig:two} (B), (C), (D)
and (E) for different noise intensities. The SR-type effect in the
bistable system is also visible, as shown in Figs.~\ref{fig:two}
(B), (C) and (D).

In Fig.~\ref{fig:two}, the bistable system, even optimized at
$D=0.8$ V$^2$/Hz via the SR effect, can not outperform the matched
filter for the case of Gaussian white noise. This fact has been
pointed repeatedly by many researchers in a variety of
investigations
\cite{Inchiosa,Bulsara&Zador,Galdi,Tougaard2002,Dykman,Peterson,Peterson&Gebeshuber,McDonnell}.
However, we have assumed complete knowledge of the probability
density functions under hypotheses $\textit{\textbf{H}}_{0}$ and
$\textit{\textbf{H}}_{1}$, from which the conclusion of the
bistable system being suboptimal by comparison with the matched
filter is deduced. In next section, we will turn to more realistic
problems in which some parameters of the signal are not completely
known. New merits of the bistable system will be observed.

\section{\label{sec:III} Evaluating the bistable system versus the matched filter
in unknown detection problems}

In this section, we mainly study two signal detection problems:
the first is detecting the pulse signal with disturbed pulse rates
$1/T_{p}$; the second is a situation that the return pulse signal
is delayed by the propagation time of the signal through the
medium, resulting in the unknown arrival time $t_0$. Both the
bistable system and the matched filter are employed as detectors
in the two signal detection problems. The performances of the
bistable system and the matched filter, with different
measurements in different detection tasks, are compared
thoroughly.

\subsection{The performance of the bistable system and the matched
filter in detecting the pulse signal with disturbed pulse rates}

Consider the heuristic problem of detecting the received pulse
signals with changed pulse rates (i.e.~multi-rate pulse signals),
for example, which may occur due to the desynchronization of
devices during the detection of moving objects. For a given noisy
environment, we can deduce the response time of the bistable
system in terms of the signal amplitude $A$, system parameters
$\tau_{a}$ and $X_{b}$ (ref. Appendix~A). If the disturbed pulse
duration $T_p$ is not too large (or small) with respect to the
system response time $T_r$, it will be seen in the following that
the bistable system can detect different pulse rates from a
frequency spectrum analysis. It is well known that intersymbol
errors will appear in this situation with a matched filter
\cite{Kay1998}. The frequency spectrum analysis of the matched
filter output also shows that too many low frequency components
degrade the matched filter's detectability.
\begin{figure}
\centering{\resizebox{6.2cm}{!}{\includegraphics{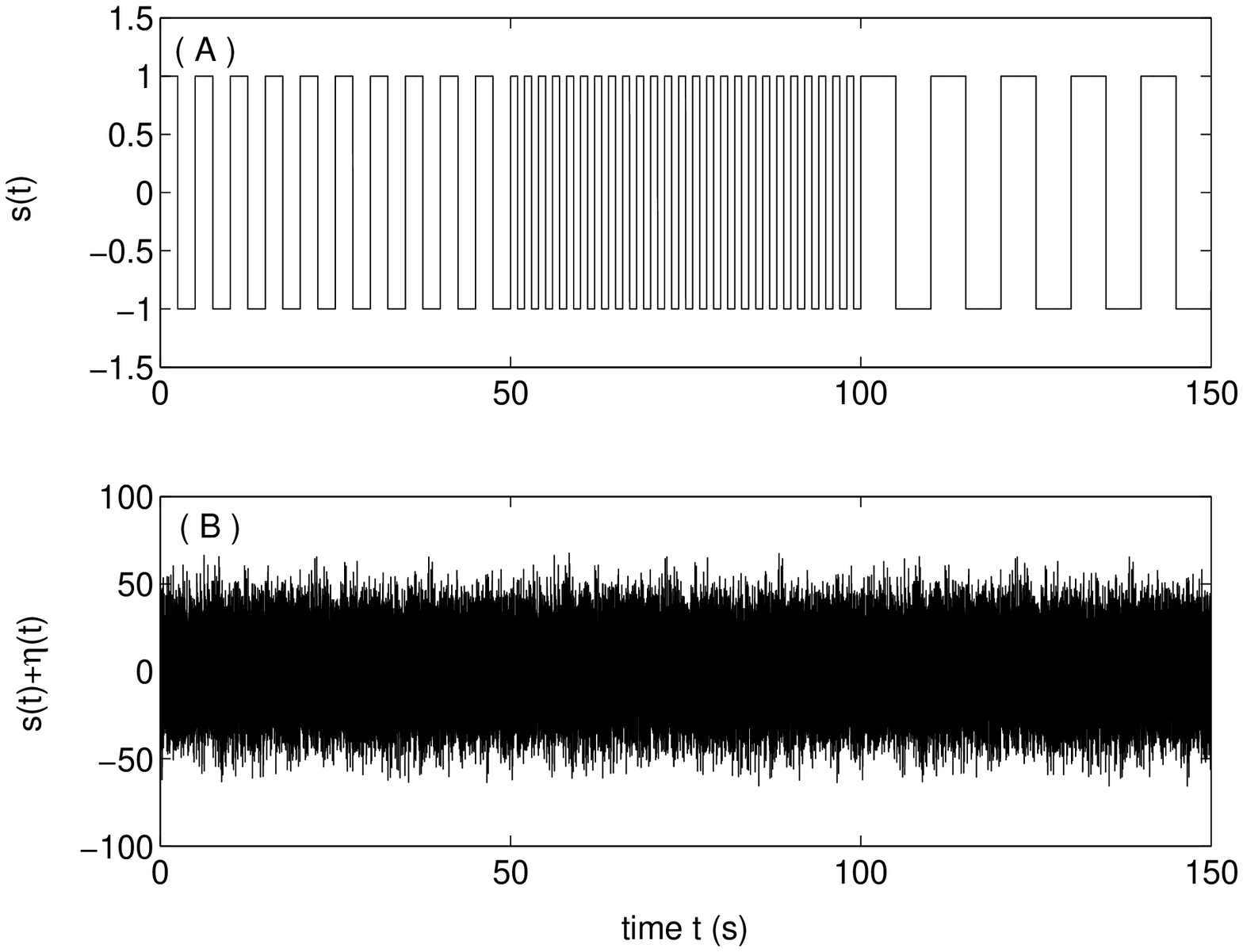}}}
\centering{\resizebox{6.2cm}{!}{\includegraphics{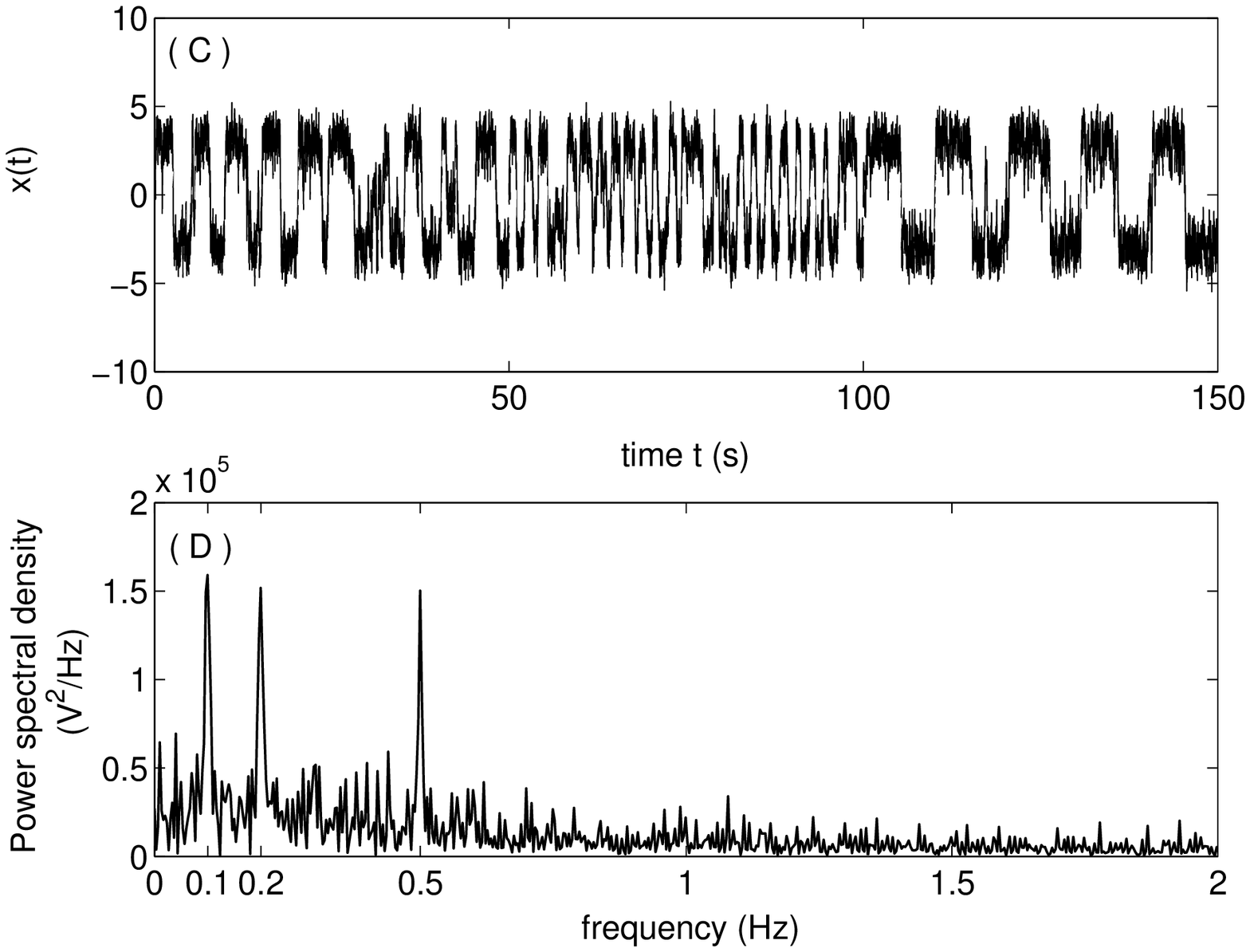}}}
\centering{\resizebox{6.2cm}{!}{\includegraphics{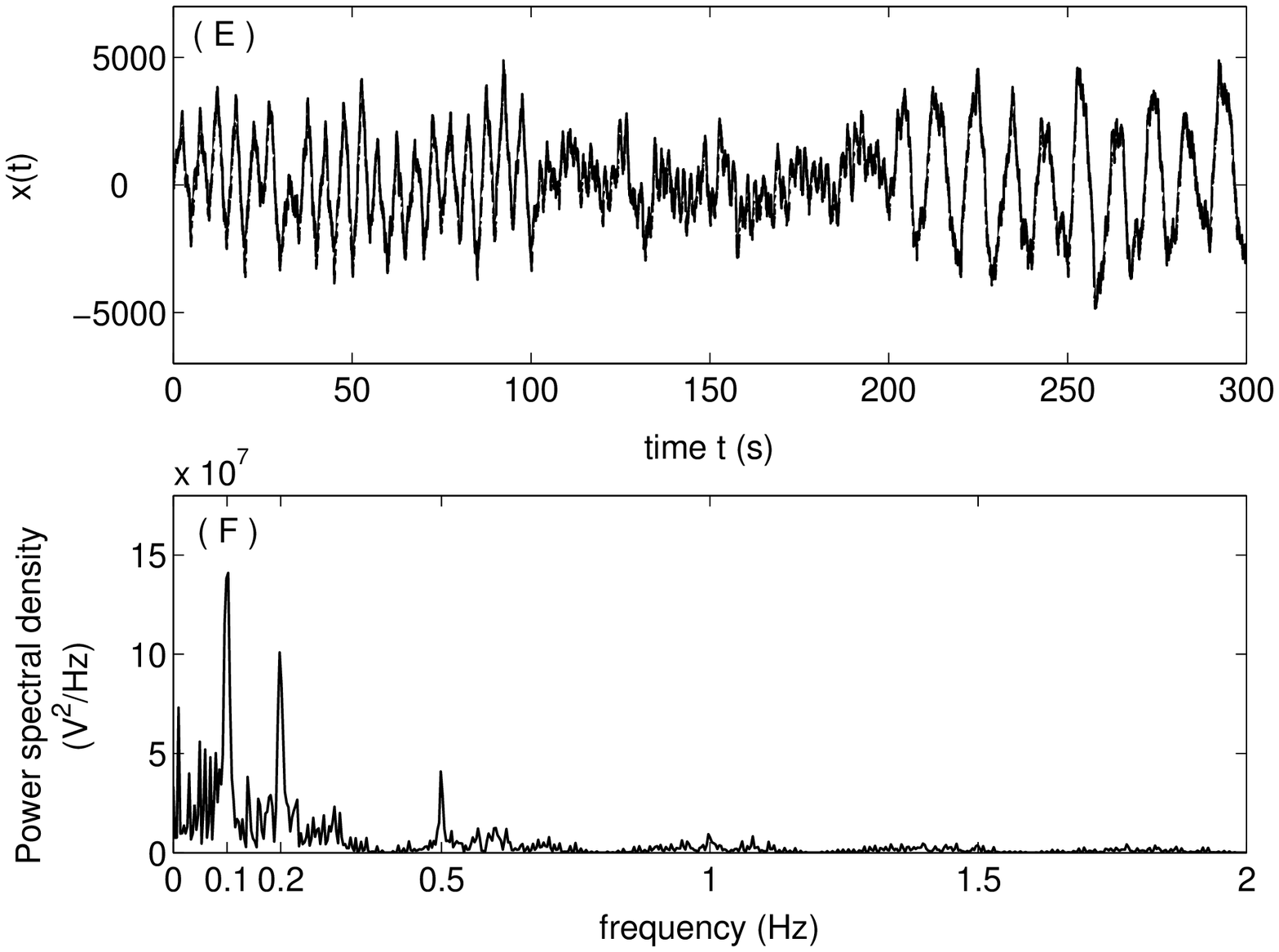}}}
\caption{(A) Input pulse signal $s(t)=\pm 1$ V with multi-rates
$1/T_p=0.2$ Hz, $0.5$ Hz and $0.1$ Hz; (B) The mixture of input
$s(t)$ plus noise $\eta(t)$ ($D=0.15$ V$^2$/Hz); (C) The output
signal $x(t)$ of the bistable system with $\tau_a=0.1$ s and
$X_b=2.8$ V; (D) The power spectrum of bistable system outputs;
(E) The output signal $x(t)$ of the matched filter. The impulse
response of $h(t)$ is selected in terms of the reference pulse
rate of $0.2$ Hz; (F) The power spectrum of matched filter
outputs. $\Delta t=0.001$ s. \label{fig:three} }
\end{figure}

In numerical simulations, the reference pulse rate is set as
$1/T_p=0.2$ Hz. Then, for the matched filter, the impulse response
should be the ``flipped around'' version of the signal, i.e.
$h(t)=s(5-t)$, $t\in[0,\: 5]$ s. We assume that the received pulse
rates of signal $s(t)$ is distributed as $0.5$ Hz and $0.1$ Hz.
Figure~\ref{fig:three} (A) shows the disturbed pulse signal $s(t)$
with three rates of $0.2$ Hz, $0.5$ Hz and $0.1$ Hz. From the
output signals of the bistable system and the matched filter, i.e.
Figs.~\ref{fig:three} (C) and (E), the corresponding power spectra
are computed and plotted in Figs.~\ref{fig:three} (D) and (F)
respectively. The power spectrum of the bistable system output, as
seen in Fig.~\ref{fig:three} (D), clearly illustrates three pulse
rates of input signal $s(t)$ at $1/T_p=0.1$ Hz, $0.2$ Hz and $0.5$
Hz. However, the frequency component at $0.5$ Hz, shown in the
power spectrum of the matched filter output, is weaker than the
spectrum lines in the low frequency domain $1/T_p<0.1$ Hz, as seen
in Fig.~\ref{fig:three} (F). This indicates that the bistable
system seems to suppress the noise at all frequency bands, but the
matched filter, more like a lowpass characteristic, allows too
much noise at low frequencies to interfere with its frequency
discrimination.

In terms of the above comparisons, we argue that the bistable
system is not so sensitive to the spread of the pulse rate values
as the matched filter. This robust feature of the bistable system
is not trivial in signal detection. Note that there exists an
optimal pulse rate matching the bistable system for fixed signal
amplitude and noise intensity \cite{Benzi1,Benzi2,Gammaitoni1998},
while the interest here is the system's robust property.
\subsection{The performance of the bistable system versus the matched
filter in detecting the pulse signal with unknown arrival time}

\begin{figure}
\centering{\resizebox{6.0cm}{!}{\includegraphics{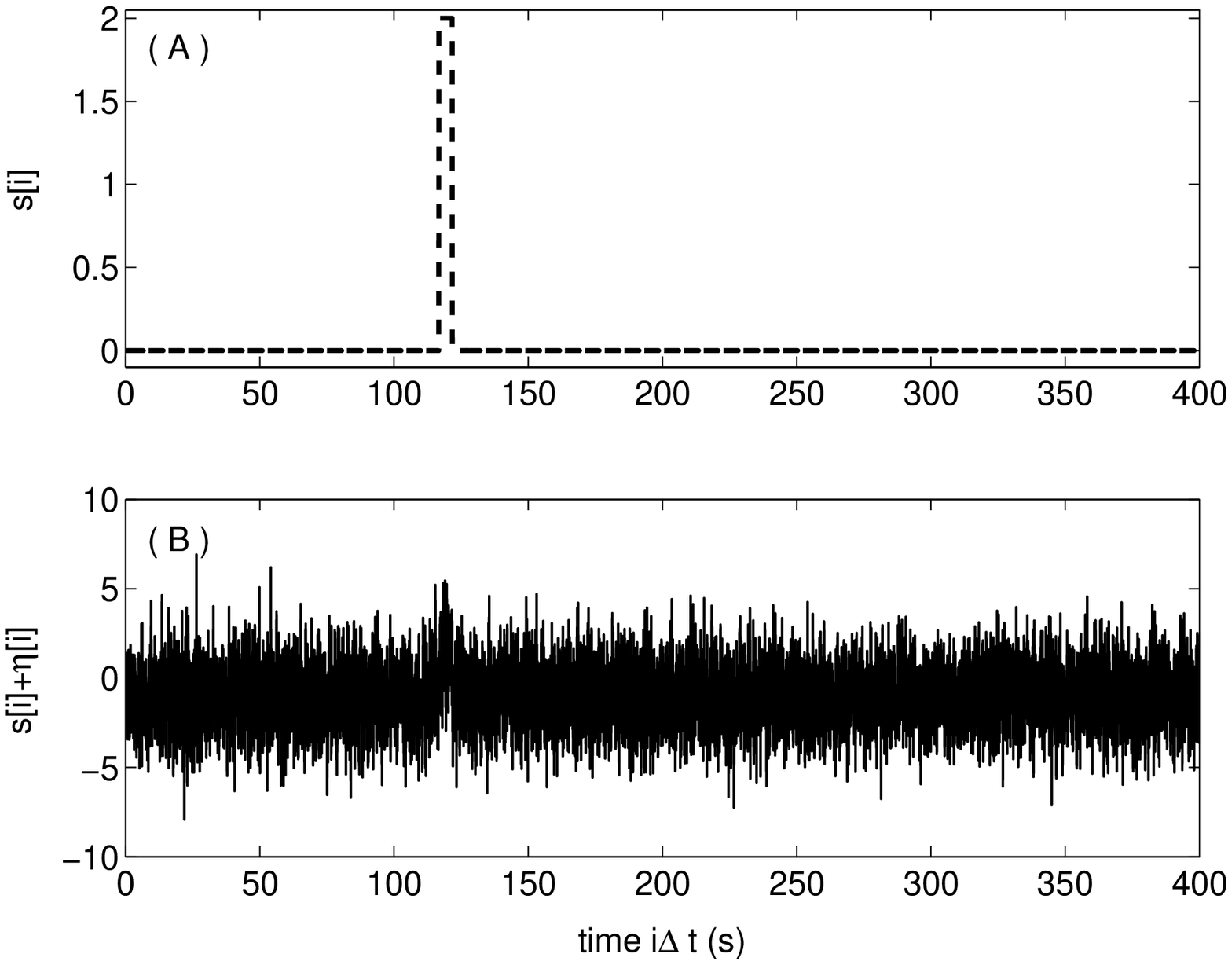}}}
\centering{\resizebox{6.2cm}{!}{\includegraphics{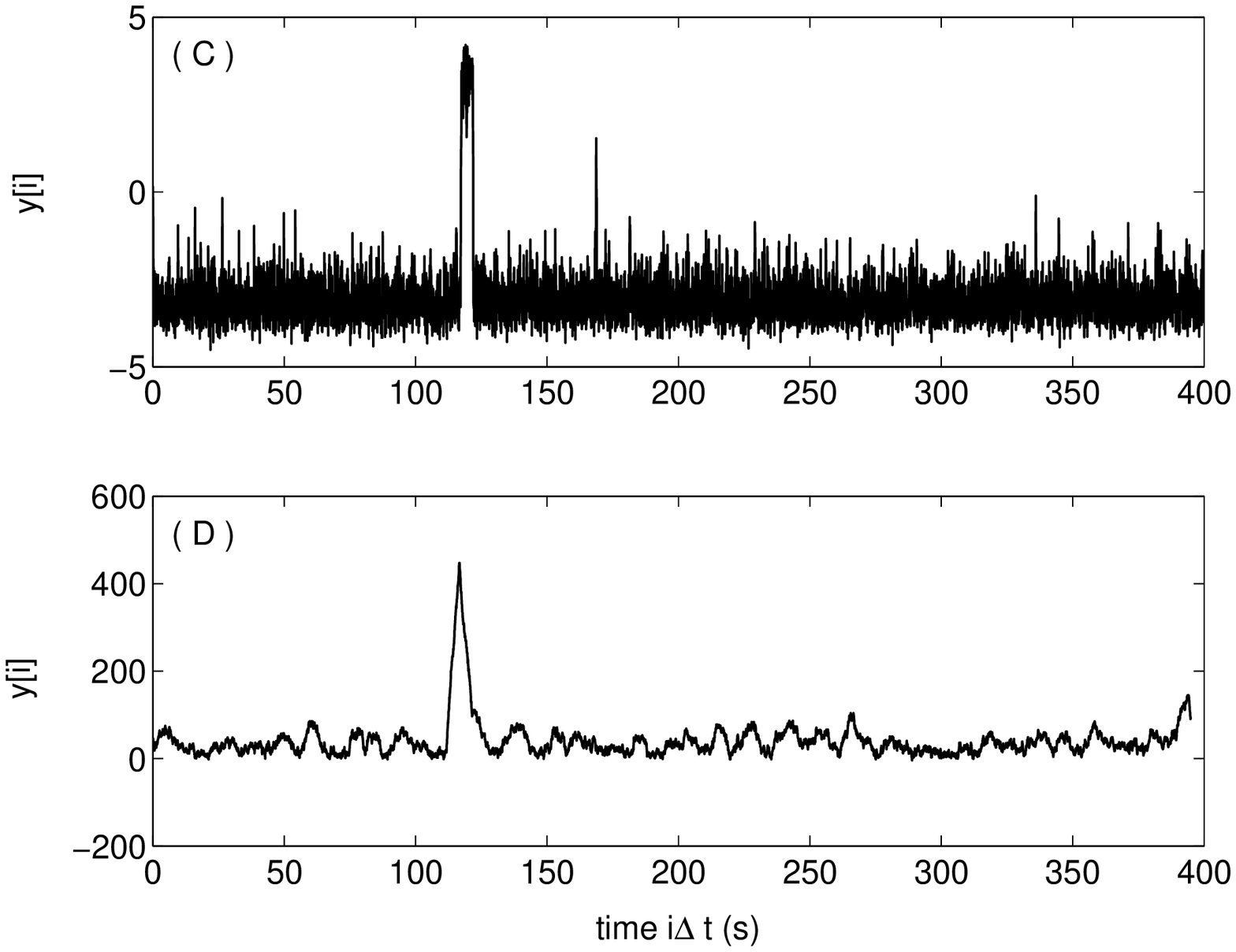}}}
\caption{ (A) Input signal $s[i]$ is one pulse with the amplitude
$2A=2$ V and the pulse duration $T_{p}=5$ s; (B) The mixture of
the signal $s[i]$ plus noise $\eta[i]$ ($D=0.08$ V$^2$/Hz); (C)
The output signal of the subthreshold bistable system with
$X_b=2.8$ V and $\tau_a=0.1$ s; (D) The convolution output signals
computed by the matched filter,
$y[i]=\sum_{i=\hat{i}_{0}}^{\hat{i}_{0}+M-1}h[i]x[i]$. $\Delta
t=0.05$ s. \label{fig:four}}
\end{figure}

In this subsection, we consider the detection problem of the
received pulse signal with an unknown arrival time $t_0$. The
hypotheses can be represented as
\begin{eqnarray}
\textit{\textbf{H}}_{0}: x[i]=\eta[i],\nonumber \\
\textit{\textbf{H}}_{1}: x[i]=s[i-i_{0}]+\eta[i], \label{eq:nine}
\end{eqnarray}
with $i=0,1,\ldots, N-1$. Here, the observation time interval is
$[0,\: T]$ and $N=T/\Delta t$. The received pulse signal $s[i]$ is
with only one pulse, as shown in Fig.~\ref{fig:four} (A). The
pulse amplitude is $2A$ over the interval $[i_0,\: i_0+M-1]$
($M=T_p/\Delta t$), but the arrival time $t_0=i_0\Delta t$ is
unknown.

For the matched filter, the test statistic can be written as
\begin{equation}
H(y)=\max_{i \in [0,\:
N-M+1]}\sum_{i=\hat{i}_{0}}^{\hat{i}_{0}+M-1}h[i]x[i]\geq
\gamma,\label{eq:ten}
\end{equation}
where $h[i]=s[M-1-i]$ for $i\in[0,$ $M-1]$ and the decision
threshold $\gamma \leq \varepsilon=\sum_{i=0}^{M-1} s^2[i]$
\cite{Kay1998}. The arrival time of input pulses is then $\hat
t_{0}=\hat{i}_{0}\Delta t +T_{p}(\varepsilon-\gamma)/\varepsilon$.
In order to find the arrival time $\hat t_0$, the matched filter
will compute $T/\Delta t$ convolutions in this detection problem.
Figure~\ref{fig:four} (D) gives an example of the convolution
output signals of the matched filter.

Considering the symmetric characteristics of bistable systems, a
bias level $-A$ is added to the input $x[i]$, i.e.,
$x[i]=s[i-i_{0}]+n[i]-A$ or $x[i]=n[i]-A$. After the bistable
system with output $y[i]$, illustrated in Fig.~\ref{fig:four} (C),
the test statistic can be expressed as
\begin{equation}
H(y)=y[\hat{i}_{0}]\geq x_+, \: \hat{i}_{0}\in [0,\: N-M+1],
\label{eq:eleven}
\end{equation}
where $x_+$ is the positive root of equation $x-x^3/X_b^2+A$
\cite{Duan2004}. Here, we assume that the bistable system needs a
time interval of $T_{r}$, i.e.~the system response time introduced
in Appendix~A, to reach the corresponding steady-state $x_+$. The
arrival time of the pulse signal is then approximated as $\hat
t_{0}=\hat{i}_{0}\Delta t-T_{r}$.
\begin{figure}
\centering{\resizebox{8cm}{!}{\includegraphics{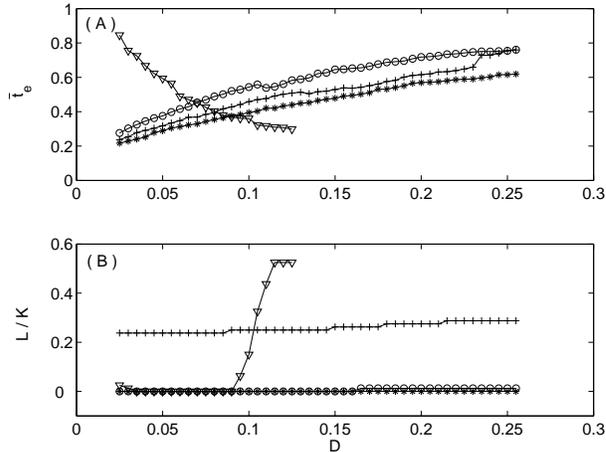}}}
\caption{(A) Average arrival time $\bar t_{e}$ versus $D$; (B)
Loss ratio $L/K$ versus $D$. For different values of $D$, each
subject is simulated with $K=80$ different arrival times $t_{0}$.
$A=1$ V, $T_p=5$ s, $T=400$ s and $\Delta t=0.05$ s. The circle
symbols represent the performance of the suprathreshold bistable
system with $X_{b}=1\;V <\sqrt{27}A/2$ and $\tau_{a}=1$ s, while
the down triangle symbols correspond to the subthreshold bistable
system with $X_{b}=2.8\; V>\sqrt{27}A/2$ and $\tau_{a}=0.1$ s. The
performance curves of the matched filter are plotted for the
decision levels $\gamma=200$ V$^2$ ($\ast$) and $400$ V$^2$ ($+$),
with $h(i)=s[99-i]$, $i \in [0,\: 99]$ and $\varepsilon=400$
V$^2$. \label{fig:five}}
\end{figure}

There are no general theories for evaluating performance of the
bistable system and the matched filter in this detection problem
\cite{Kay1998}. In numerical simulations, the performance of both
detectors is quantified by the average error time
\begin{equation}
\bar t_{e}=\sum_{j=1}^{K}\frac{\mid \hat t_{0}[j]-t_{0} \mid}{K},
\label{eq:twelve}
\end{equation}
where $K$ is the simulation time and $j=1,2,\ldots,K$. Here, $\hat
t_{0}[j]$ is the $j_{\rm th}$ arrival time of the signal given by
the detector. The matched filter or the bistable system, however,
might detect a totally wrong arrival time $\hat t_{0}$, which is
measured by the loss ratio $L/K$ and the loss time $L$. If we take
$\mid \hat t_{0}[m]-t_{0} \mid
>T_{p}$ ($m=1,2,\cdots,L$) as one loss time, the average error
time $\bar t_{e}$ can be rewritten as
\begin{equation}
\bar t_{e}=\sum_{j=1}^{K-L}\frac{\mid \hat t_{0}[j]-t_{0}
\mid}{K-L}.\label{eq:thriteen}
\end{equation}
Moreover, we know the computation time $T_c[j]$ of the bistable
system and matched filter in the $j_{\rm th}$ numerical
simulation. Thus, the computational efficiency of both detectors
can be quantified by the average computation time
\begin{equation}
\bar T_{c}=\sum_{j=1}^{K}\frac{T_{c}[j]}{K}.\label{eq:fourteen}
\end{equation}

The plots of Figs.~\ref{fig:five}~(A) and (B) show the performance
of bistable systems and the matched filter in detecting the
unknown arrival time of input signals, this is, the average error
time $\bar t_{e}$ and the loss ratio $L/K$. It is seen that the
matched filter's performance depends on selecting an appropriate
decision threshold $\gamma$ \cite{Kay1998}. If the decision
threshold $\gamma=400$ V$^2$, it will present a higher loss ratio
$L/K$ than in bistable systems. The decision threshold
$\gamma=200$ V$^2$ makes the matched filter optimized. Two kinds
of bistable systems are employed in this detection problem: one is
subthreshold ($X_b>\sqrt{27}A/2$) and the other is suprathreshold
($X_b<\sqrt{27}A/2$) \cite{Moss,Duan2004,Xu2004}. In
Fig.~\ref{fig:five}, the average error time $\bar t_{e}$ and the
loss ratio $L/K$ of the suprathreshold bistable system is a
monotonic increasing function of the noise intensity $D$. This
suprathreshold bistable system is suboptimal compared to the
matched filter with $\gamma=200$ V$^2$, but outperforms the
matched filter with $\gamma=400$ V$^2$. For the subthreshold
bistable system shown in Fig.~\ref{fig:five}, the average error
time $\bar t_{e}$ decreases as the noise intensity $D$ increases,
and almost approaches the performance of the matched filter
($\gamma=200$ V$^2$) before the loss ratio $L/K$ begins to
increase. This non-monotonic influence of the noise on the
performance of the subthreshold bistable system is the signature
of the SR phenomenon.

Moreover, an important merit of the bistable system is its
computational efficiency. The average computation time of the
bistable system, i.e.~Eq.~(\ref{eq:fourteen}), is about $\bar
T_{c}=0.3361$ s for processing the data observed in the time
interval $[0,\: T]$ ($T=400$ s). The matched filter will use about
$\bar T_{c}=0.9048$ s to deal with the same data. This
computational characteristic of the bistable system maybe
meaningful in the optimization of speed and efficacy of
information processing with limited resources for data handling,
storage and energy supply
\cite{Galdi,Chapeau-Blondeau2002,Papadopoulos1998}.

\section{Conclusion}

We thoroughly evaluated the performance of the nonlinear bistable
system versus the linear matched filter in pulse signal detection
problems. Based on an approximate theory of the probability
density model, the signal detection statistics of the bistable
system were deduced theoretically. The detection performance of
the nonlinear bistable system can be improved by adding an optimal
amount of noise, i.e.~the SR phenomenon, which was also
demonstrated by numerical experiments. The theoretical and
numerical results here lead to the fact, emphasized by many
researchers
\cite{Inchiosa,Bulsara&Zador,Galdi,Tougaard2002,Dykman,Peterson,Peterson&Gebeshuber,McDonnell},
that the bistable system is suboptimal to the matched filter for
detecting the signal with complete known parameters. However, the
bistable system is superior in detecting a pulse signal with
unknown parameters. The power spectrum analysis demonstrated that
the bistable system is more robust than the matched filter for
discriminating the disturbed pulse rate. In the case of detecting
the unknown arrival time of the pulse signal, the bistable system
is also comparable to the matched filter. Furthermore, the
detection scheme of the bistable system is more efficient with
computational saving. These significant results verified the
potential applicability of nonlinear bistable systems in the
signal detection field, which might deserve further studies, for
example, in exploiting other nonlinear systems exhibiting SR
effects in practical detection problems.
\section*{Acknowledgements}
This project is sponsored by Scientific Foundation for Returned
Overseas Chinese Scholars, Ministry of Education, and the Natural
Science Foundation of Shandong Province of P. R. China (No.
Y2002G01). Funding from the Australian Research Council (ARC) is
gratefully acknowledged.

\appendix
\section{System response time and nonstationary probability density model}
\subsection{System response time} In Eq.~(\ref{eq:three}), the
Fokker-Planck operator $L_{FP}=\frac{\partial}{\partial
x}V'(x)+\frac{D}{\tau_a}\frac{\partial^2}{\partial x^2}$ is not a
Hermitian operator \cite{Risken}. We rescale the variables as
\begin{equation}
\bar X_b=X_b/ \sqrt{D/\tau_a}, \: \bar A =A/ \sqrt{D/\tau_a}, \:
\tau=t/ \tau_a,\: y=x/ \sqrt{D/\tau_a}, \label{eq:A1}
\end{equation}
Eq.~(\ref{eq:three}) becomes
\begin{equation}
\frac{\partial\rho(y,\tau)}{\partial
\tau}=\big[\frac{\partial}{\partial
y}V'(y)+\frac{\partial^2}{\partial y^2}\big] \rho(y,\tau),
\label{eq:A2}
\end{equation}
where $V'(y)=-y+y^3/\bar X_b^2\mp \bar A$. The steady-state
solution of Eq.~(\ref{eq:A2}) is given by
\begin{equation}
\rho(y)=\lim_{\tau\rightarrow\infty}\rho(y,\tau)=C\exp[-V(y)],\label{eq:A3}
\end{equation}
where $C$ is the normalization constant. A separation ansatz for
$\rho(y,\tau)$ \cite{Risken},
\begin{equation}
\rho(y,\tau)=u(y)\exp[-\frac{V(y)}{2}]\exp(-\lambda \tau),
\label{eq:A4}
\end{equation}
leads to
\begin{equation}
L u=-\lambda u, \label{eq:A5}
\end{equation}
with a Hermitian operator $L=(\partial^2/\partial
y^2)-[\frac{1}{4}V'^2(y)-\frac{1}{2}V''(y)]$. The functions $u(y)$
are eigenfunctions of the operator $L$ with the eigenvalues
$\lambda$. Multiplying both sides of Eq.~(\ref{eq:A5}) by $u(y)$
and integrating it, yields
\begin{equation}
\lambda=\frac{\int_{-\infty}^{+\infty}\{u'^2(y)+u^2(y)[\frac{1}{4}V'^2(y)-\frac{1}{2}V''(y)]\}dy}
{\int_{-\infty}^{+\infty}u^2(y)dy},\label{eq:A6}
\end{equation}
where eigenfunctions $u(y)$ satisfy the boundary conditions of
$\lim_{y \to \pm\infty}u(y) =0$ and $\lim_{y \to
\pm\infty}u'(y)=0$. The eigenvalue problem of Eq.~(\ref{eq:A5}) is
then equivalent to the variational problem consisting in finding
the extremal values of the right side of Eq.~(\ref{eq:A6}). The
minimum of this expression is then the lowest eigenvalue
$\lambda_0=0$, corresponding to the steady state solution of
Eq.~(\ref{eq:A3}) \cite{Risken}. Here, we adopt eigenfunctions
$u(y)=p(y)\exp[-V(y)/2]$ and $p(y)\neq 0$, Eq.~(\ref{eq:A6}) then
becomes
\begin{equation}
\lambda=\frac{\int_{-\infty}^{+\infty}\{p'^2(y)+\frac{1}{2}p^2(y)V'^2(y)
-\frac{1}{2}[V'(y)p^2(y)]'\}\exp[-V(y)]dy}{\int_{-\infty}^{+\infty}p^2(y)\exp[-V(y)]dy}.\label{eq:A7}
\end{equation}
Since
\begin{eqnarray*}
\int_{-\infty}^{+\infty}[V'(y)p^2(y)]'\exp[-V(y)]dy=
V'(y)p^2(y)\exp[-V(y)]|_{-\infty}^{+\infty}\\
+\int_{-\infty}^{+\infty}p^2(y)V'^2(y)\exp[-V(y)]dy=
\int_{-\infty}^{+\infty}p^2(y)V'^2(y)\exp[-V(y)]dy,
\end{eqnarray*}
Eq.~(\ref{eq:A7}) can be rewritten as
\begin{equation}
\lambda=\frac{\int_{-\infty}^{+\infty}p'^2(y)\exp[-V(y)]dy}
{\int_{-\infty}^{+\infty}p^2(y) \exp[-V(y)]dy}. \label{eq:A8}
\end{equation}
Assume $p(y)=d_0+d_1 y+\cdots+d_n y^n$ and the order $n$ is an
integer, we obtain
\begin{equation}
([K]-\lambda [M])\{d\}=0, \label{eq:A9}
\end{equation}
with eigenvectors $\{d^i\}=[d_0^i, d_1^i, \ldots, d_n^i]$
corresponding to eigenvalues $\lambda_i$ for $i=0,1,\ldots,n$. The
elements of matrices $[M]$ and $[K]$ are
\begin{eqnarray*}
m_{ij}=\int_{-\infty}^{+\infty}y^{i+j}\exp[-V(y)]dy>0,\\
k_{ij}=\int_{-\infty}^{+\infty} ij y^{i+j-2}\exp[-V(y)]dy\geq0,
\end{eqnarray*}
where $i,$ $j=0,1,\ldots,n$. The matrix $[M]$ is positive definite
and the matrix $[K]$ is semi-positive definite. The minimal
eigenvalue $\lambda_0$ is zero. The inverse of minimal positive
eigenvalue $\lambda_1$ describes the main time of the system
tending to the steady state solution of Eq.~(\ref{eq:A3}), what we
call the system response time. Note the time scale transformation
in Eq.~(\ref{eq:A1}), the minimal positive eigenvalue should be
$\lambda_1/\tau_a$ and the real system response time is
$T_r=\tau_a/\lambda_1$.

\subsection{Nonstationary probability density model}

From Eq.~(\ref{eq:A9}), we can obtain the eigenfunctions
$u_i(y)=p_i(y)\exp[-V(y)/2]$ corresponding to the eigenvalue
$\lambda_i$ for $i=0,1,\ldots,n$, where $p_i(y)=d_0^i+d_1^i
y+\cdots+d_n^i y^n$. The eigenvectors $\{d^i\}=[d_0^i,
d_1^i,\ldots,d_n^i]$ are normalized. Because $L$ is a Hermitian
operator, eigenfunctions $u_i(y)$ and $u_j(y)$ are orthogonal
\begin{equation}
\int_{-\infty}^{+\infty}u_i(y)u_j(y)dy=\delta_{ij}, \label{eq:A10}
\end{equation}
where $i,j=0,1,\ldots,n$. $\rho(y,\tau)$ can be expanded,
according to eigenfunctions $u_i(y)$ and eigenvalues $\lambda_i$,
as
\begin{equation}
\rho(y,\tau)=\sum_{i=0}^n C_i u_i(y)\exp[-\frac{V(y)}{2}]
\exp[-\lambda_i\tau], \label{eq:A11}
\end{equation}
where $C_{i}$ are normalization constants. In this paper, we take
an approximate expression of $\rho(y,\tau)\simeq\sum_{i=0}^1 C_i
u_i(y)\exp[-V(y)/2] \exp[-\lambda_i\tau]$ instead of
Eq.~(\ref{eq:A11}). Then, if the preceding input signal is
$s(t)=\mp A$ and the next one is $s(t)=\pm A$, a simple
nonstationary probability density model is derived as
\begin{equation}
\rho(x,t|s(t)=\pm A)\simeq \rho(x|s(t)=\pm A) +[\rho(x|s(t)=\mp
A)-\rho(x|s(t)=\pm A)]\exp(-t/T_r), \label{eq:A12}
\end{equation}
with the initial condition $\rho(x,t=0|s(t)=\pm A)=\rho(x|s(t)=\mp
A)$ and the stationary condition $\rho(x,t=+\infty|s(t)=\pm
A)=\rho(x|s(t)=\pm A)$, respectively. Here, $\rho(x|s(t)=\pm A)$
are the steady-state solutions given in Eq.~(\ref{eq:four}), and
$V(x)=-x^2/2+x^4/(4X_b^2)\mp Ax$ correspond to the constant inputs
$s(t)=\pm A$ respectively. This kind of nonstationary solution of
$\rho(x,t|s(t)=\pm A)$ can be further developed into the case of
$\rho(y,\tau)\simeq\sum_{i=0}^n C_i u_i(y)\exp[-V(y)/2]
\exp[-\lambda_i \tau]$ for $n \geq 2$.

\end{document}